\documentclass [amssymb, amsmath, showpacs,nofootinbib]{revtex4}

\usepackage{graphicx}
\usepackage{dcolumn}
\usepackage{bm}
\usepackage[all]{xy}

\begin{document}

\title{Fermions and Loops on Graphs. II. Monomer-Dimer Model as Series of Determinants.}
\author {Vladimir Y. \surname{Chernyak}$^{a,b}$}
\author{Michael \surname{Chertkov}$^{b}$}

\affiliation{$^a$Department of Chemistry, Wayne State University,
5101 Cass Ave,Detroit, MI 48202}

\affiliation{$^b$Center for Nonlinear Studies and Theoretical Division, LANL, Los Alamos, NM 87545}

\date{\today}

\begin{abstract}
We continue the discussion of the fermion models on graphs that started in the first paper of the
series. Here we introduce a Graphical Gauge Model (GGM) and show that : (a) it can be stated as an
average/sum of a determinant defined on the graph over $\mathbb{Z}_{2}$ (binary) gauge field; (b)
it is equivalent to the Monomer-Dimer (MD) model on the graph; (c)  the partition function of the
model allows an explicit expression in terms of a series over disjoint directed cycles, where each
term is a product of local contributions along the cycle and the determinant of a matrix defined
on the remainder of the graph (excluding the cycle). We also establish a relation between the MD
model on the graph and the determinant series, discussed in the first paper, however, considered
using simple non-Belief-Propagation choice of the gauge. We conclude with a discussion of possible
analytic and algorithmic consequences of these results, as well as related questions and
challenges.
\end{abstract}

\pacs{02.50.Tt, 64.60.Cn, 05.50.+q}

\maketitle

Gauge theories, stated in terms of fermions and gauge fields (e.g. associated with a vector
potential), are common in theoretical and mathematical physics \cite{87Pol,89Zin}. Normally in
physics, e.g. discussing quantum electrodynamics or quantum gravity, these popular theories are
defined over continuous spaces, or their natural discretizations, e.g. triangulated Eucledian
grids. In the discretized versions, e.g. in Lattice Gauge Theories \cite{74WK}, fermions are
normally associated with vertexes of the grid, while gauge variables reside on edges.

In this paper we extend this standard discretized construction to arbitrary graphs and show that
the gauge theory approach, native of physics, can be useful for getting new nontrivial relations
between different graphical models that describe computer science problems defined on arbitrary
graphs. We introduce and discuss a Graphical Gauge Model (GGM). Gauge fields in our construction
correspond to standard binary variables, which could also be called Ising spins/variables, or more
formally, the gauge group of the theory is $\mathbb{Z}_{2}$. Two objects emerging in any
gauge theory, determinants and loops, are therefore natural participants of our description. We
also find that this approach and language fits naturally with the Loop Calculus introduced in
\cite{06CCa,06CCb} and  extended in the first paper of the series \cite{08CCa} to the Gaussian
Graphical models on graphs.

The power of GGM is in its natural operational flexibility: changing the order of integrations and
modifying the integrand in the expression for the model's partition function result in a variety of
nontrivial relations, some of them discussed in this paper. Integration over the Grassman-fermion
variables turns the partition function of GGM into a $\zeta$-function dependent on the gauge
(binary variable) configuration. Here the $\zeta$-function is understood as a generating function
for the expectation values of the Grassman variables and their combinations. In this formulation it
is related to the inverse of the Ihara $\zeta$-function of the graph \cite{66Iha}.

Even though we are making a point in promoting the language of Grassman/fermion integration in this
paper, and the series in general, our two main statements are made in terms of ``normal'' objects,
e.g. determinants, disjoint oriented cycles, and also partition functions of the monomer-dimer
model. Given that this latter object did not appear in the first paper of the series, we find it
useful to state it casually right away (see the rhs of Eq.~(\ref{MD2}) for formal definition).
Consider a graph, and cost functions, $w_a$ and $w_{ab}$, associated with the vertices and edges of
the graph. A monomer-dimer configuration on the graph is a set of colorings of vertices and edges
so that either any vertex of the graph is colored and then no adjusted edges are colored, or the
vertex is not colored but then one of the adjusted edges is colored. The partition function of the
monomer-dimer model on the graph is the sum over all allowed monomer-dimer
configurations/colorings,  where each individual contribution is a product of factors associated
with colored vertices and edges over the graph \footnote{Note, that ``dimers" and ``monomers" are
terms used in statistical physics which are also fully equivalent to ``perfect matchings" and
``imperfect matchings" in the terminology commonly accepted in computer science, see e.g.
\cite{86LP}.}. Armed with this definition let us now state the main results reported in the paper:
\begin{itemize}

\item  The partition function, $Z_{MD}$, of the monomer-dimer model on a
graph ${\cal G}$ is expressed in terms of a matrix $H$  built from the monomer and dimer weights
placed at the diagonal and off-diagonal elements, respectively. Specifically, $Z_{MD}({\cal G})$ is
stated as a series over the oriented disjoint cycles $C\in {\rm ODC}({\cal G})$ of the graph. An
oriented disjoint cycle $C\in {\rm ODC}({\cal G})$ is represented by a disjoint union of simple
oriented loops. Each term in the series is equal to the determinant of the original matrix $H$ with
the cycles excluded, $H|_{{\cal G}\setminus C}$, multiplied by the product along the simple loops
of the cycle of the corresponding off-diagonal elements taken with the reversed signs. (See
Eqs.~(\ref{MD2},\ref{DCS-explicit}).)

\item  The determinant of $H$ is stated as a series over oriented disjoint
cycles $C\in {\rm ODC}({\cal G})$ of the graph ${\cal G}$, where each term is equal to the
partition function, $Z_{MD}({\cal G}\setminus C)$, of the monomer-dimer of the original graph with
the cycle excluded, multiplied by the product along the simple loops of the cycle of the
off-diagonal elements taken with the reversed signs. (See Eqs.~(\ref{LS-SL-explicit-2},\ref{MD2}).)

\end{itemize}
Three remarks are in order. First, the two main statements are equivalent, in fact, one is a kind
of an inverse of the other (see Appendix \ref{sec:app} for clarifications). Second, the first
statement has an immediate algorithmic consequence: it may be used for an approximate computation
of the monomer-dimer partition function (which is known to be an $\#P$-complete, i.e. a counting
problem of the likely exponential complexity \cite{90Jer}) via a truncation of the determinant
series (the complexity of a determinant evaluation is cubic in the graph size). Third, the
statement number two (expansion of the determinant in a series over oriented disjoint cycles) can
be derived by implementing the Gauge fixing approach in the spirit of \cite{06CCa,06CCb}, however
selecting a gauge different from the Belief Propagation (BP) gauge. The latter resulted in the Loop
Series expansion for the determinant, described in the first paper of the series \cite{08CCa}.

\begin{figure*}
\centering
\vspace{0.1in}
\includegraphics[width=0.9\textwidth]{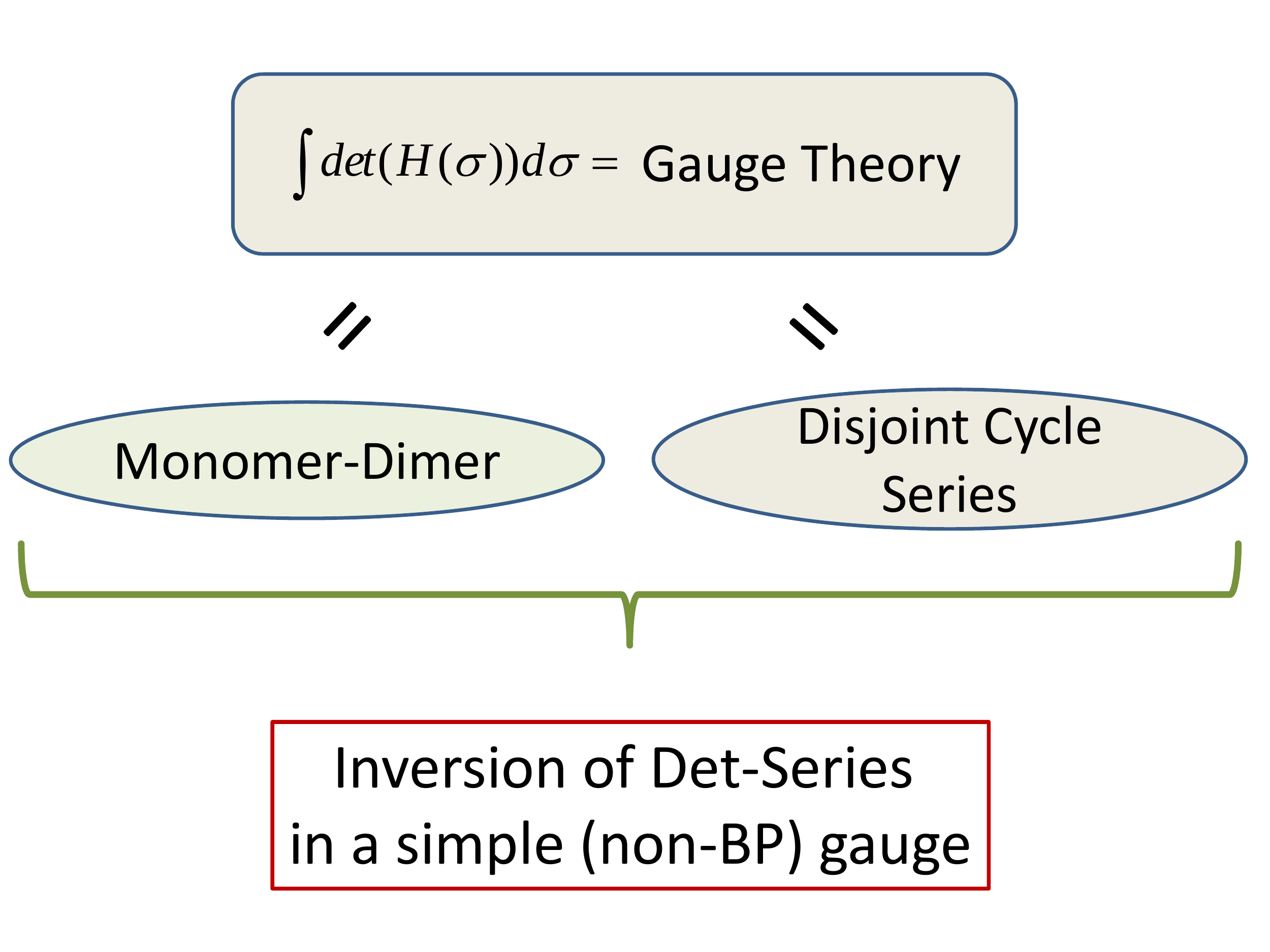}
\caption{Schematic description of relations between models and concepts discussed in the paper. }
\label{fig:scheme}
\end{figure*}

A schematic set of relations between the two main statements and other results and models discussed
in the paper, are shown in Fig.~\ref{fig:scheme}. The distribution of material is as follows. GGM
is introduced in Section \ref{subsec:GGM}. Direct relations between GGM and partition function of
MD and DCS over determinants are established in Section \ref{subsec:MD} and Section
\ref{subsec:DCS} respectively. Inverse of the relation,  expressing determinant as a series over
partition functions of MD models on the original graph with disjoint cycles excluded is discussed
in Section \ref{sec:DetMD}, with some auxiliary material placed in Appendix \ref{sec:app} and
Appendix \ref{sec:app-B}. Section \ref{sec:sum} is reserved for Summary and Conclusions.

\section{Graphical Gauge Model, Monomer-Dimer and Disjoint Cycle Series}
\label{sec:main}

\subsection{Graphical Gauge Model}
\label{subsec:GGM}

The determinant of a matrix was the key object discussed in \cite{08CCa}. Thus, we naturally start
a technical description in this second paper with stating a new model in terms of determinants.

Consider a square matrix, $H$, with elements $H_{ab}$, $a,b=1,\cdots,N$, and define a set of
transformed (twisted) matrices, $H(\bm\sigma)$, determined by a set of fields,
${\bm\sigma}=(\sigma_{ab}=\pm 1|a\neq b; \; a,b=1,\cdots,N)$, hereafter referred to as gauge
fields, according to the following rule: $H_{ab}({\bm\sigma})=\sigma_{ab}H_{ab}$ and
$H_{aa}({\bm\sigma})=H_{aa}$. The generalized $\zeta$-function of the matrix (understood as a
generating function for Grassman variables correlations) is defined simply as the determinant,
\begin{eqnarray}
 \zeta({\bm\sigma})\equiv\det\left(H({\bm\sigma})\right).
\label{zeta}
\end{eqnarray}
Note that the Ihara $\zeta$-function of a graph depends on a spectral parameter represented by a
complex number $\omega$. We add an additional set of binary spectral parameters, represented by the
gauge field components and set $\omega=0$.

The matrix $H$ also defines an undirected graph ${\cal G}(H)$ with $N$ nodes $a\in{\cal G}_{0}$.
The nodes $a$ and $b$ are connected by an edge $\alpha=\{a,b\}\in{\cal G}_{1}$ when $H_{ab}\ne 0$
or $H_{ba}\neq 0$. In other words the nodes $a\in{\cal G}_{0}$ represent the diagonal elements
$H_{aa}$, whereas the edge $\{a,b\}\in{\cal G}_{1}$ corresponds to the pair of the off diagonal
matrix elements $H_{ab}$ and $H_{ab}$ where at least one of the elements is nonzero. Hereafter we
will also use a convenient notation $a\sim b$ for $\{a,b\}\in{\cal G}_{1}$. Note that for defining
the $\zeta$-function we need only those components $\sigma_{ab}$ that are related to the edges of
${\cal G}$, i.e., $a\sim b$. Therefore, hereafter the gauge fields will include the relevant
components, only, i.e. ${\bm\sigma}=(\sigma_{ab}=\pm 1|a\neq b; \; a\sim b; \; a,b=1,\cdots,N)$.

The last comment allows the configurations ${\bm\sigma}$ to be interpreted as discrete gauge fields
with the gauge group $\mathbb{Z}_{2}$ that reside on the graph ${\cal G}$ in the full accordance
with the terminology of the Lattice Gauge Theories \cite{74WK,87Pol,89Zin}, originally developed in
the context of regular lattices. Recall that in Lattice Gauge Theories the components $\sigma_{ab}$
of a gauge field reside on the edges of the lattice and take values in the gauge group; the latter
means that in our case the gauge group is $\mathbb{Z}_{2}$. It is important to note that generally
$\sigma_{ab}\ne\sigma_{ba}$, they rather satisfy the constraints $\sigma_{ba}\sigma_{ab}=1$ (with
$1$ being naturally the gauge group unit element). However, in our special case of the
$\mathbb{Z}_{2}$ gauge group the constraints imply $\sigma_{ba}=\sigma_{ab}$, and we can interpret
the gauge field components as residing on the graph unoriented edges $\{a,b\}=\{b,a\}$, not on
plaquettes as common in standard Lattice Gauge Theories considered on surface graphs \footnote{A
general graph does not have a notion of plaquettes, and therefore the gauge field curvature
(intensity) that resides on the plaquettes may not be introduced. However, equipping the graph with
an additional structure, namely a cyclic ordering of the edges attached to a vertex for all
vertices $a\in{\cal G}_{0}$, which turns the graph into a so-called fatgraph \cite{93MP}, allows to
interpret ${\cal G}$ as the $1$-skeleton of a $2$-dimensional CW-complex that represents a Riemann
surface, where the CW-complex \cite{89Spa} is a space that can be obtained step-by-step via
attaching cells of higher dimension. The set of points constitutes its $0$-dimensional skeleton.
Attaching $1$-dimensional cells represented by the edges results in an un-oriented graph that
constitutes the $1$-dimensional skeleton. Attaching $2$-dimensional cells represented by plaquettes
results in the $2$-dimensional skeleton. In our case the latter reproduces a Riemann surface and no
more cells are attached. The gauge field intensity that resides in the $2$-dimensional cells of the
obtained CW-complex plays an important role in this case.
}.

The $\mathbb{Z}_{2}$ gauge theory associated with matrix $H$ and the graph ${\cal G}(H)$,
respectively, is stated simply as an average/sum of the gauge-field-dependent determinant over all
possible configurations of the gauge field on the graph. The partition function of the model
becomes
\begin{eqnarray}
Z=2^{-|{\cal G}_1|}\sum_{{\bm\sigma}\in{\cal G}_1}\zeta({\bm\sigma})\equiv
\int_{{\cal G}_1}{\cal D}{\bm\sigma}\det(H({\bm\sigma})),
\label{GGM1}
\end{eqnarray}
where the ``integral" over the set of nonzero edges on the rhs is simply a convenient notation for
the sum over $2^{|{\cal G}_1|}$ possible states of the discrete gauge fields and $|{\cal G}_1|$
stands for the cardinality of ${\cal G}_1$ (i.e., number of edges of ${\cal G}$).

Obviously one can think of any determinant on the rhs of Eq.~(\ref{GGM1}) as of the one derived in
the result of averaging/integration over Grassman variables ${\bm \theta}$, associated with the
vertexes of ${\cal G}$. Adopting the notation introduced in the first paper \cite{08CCa} (see also
\cite{87Ber}) we can recast the partition function (\ref{GGM1}) in a form
\begin{eqnarray}
&& Z=\int_{{\cal G}_1}{\cal D}{\bm\sigma}\int {\cal D}{\bm\theta}{\cal D}\bar{\bm\theta}
\exp\left({\cal S}_0(\bar{\bm\theta},{\bm\theta};{\bm\sigma})\right),
\label{GGM2}\\
&&  S_{0}(\bar{{\bm\theta}},{\bm\theta};{\bm\sigma})=\sum_{a\in{\cal
G}_{0}}H_{aa}\bar{\theta}_{a}\theta_{a}+\sum_{\{a,b\}\in{\cal
G}_{1}}\sigma_{ab}(H_{ab}\bar{\theta}_{a}\theta_{b}+H_{ba}\bar{\theta}_{b}\theta_{a}).
\label{S0}
\end{eqnarray}
Following the terminology commonly accepted in the field theory and mathematical physics we call
$S_0$ the action of the Graphical Gauge Model. Note that, since the action in Eq.~(\ref{S0})
depends on the gauge field ${\bm\sigma}$, it describes free fermions, interacting with the gauge
field. The action of the pure gauge field in this model is zero.

\subsection{Monomer-Dimer Model}
\label{subsec:MD}

The integrations/summations on the rhs of Eq.~(\ref{GGM2}) obviously commutes,  thus exchanging the
order of integration, expanding vertex terms of the integrand in the series, utilizing the
anti-commuting features of the Grassman variables and, finally, integrating over the binary gauge
variables, one derives
\begin{eqnarray}
\label{MD1} \int_{{\cal G}_1}{\cal
D}{\bm\sigma}e^{S_{0}(\bar{{\bm\theta}},{\bm\theta};{\bm\sigma})}=\prod_{a\in{\cal
G}_{0}}(1+w_{a}\bar{\theta}_{a}\theta_{a})\prod_{\{a,b\}\in{\cal
G}_{1}}(1+w_{ab}\bar{\theta}_{a}\theta_{a}\bar{\theta}_{b}\theta_{b}),
\end{eqnarray}
where $w_{ab}\equiv-H_{ab}H_{ba}$ and  $w_a\equiv H_{aa}$. Expanding Eq.~(\ref{MD1}) into a
polynomial and integrating the resulting expression over the Grassman variables we find that only
terms associated with valid monomer-dimer configurations survive (are nonzero), i.e.
\begin{eqnarray}
Z=Z_{MD}\equiv \sum_{\bm\pi}\left(\prod_{a\in{\cal G}_0} w_a^{\pi_a}\right)
\left(\prod_{\{a,b\}\in{\cal G}_1}w_{ab}^{\pi_{ab}}\right) \left(\prod_{a\in{\cal G}_0}
\delta\left(\pi_a+\sum_{b\sim a}\pi_{ab},1\right)\right), \label{MD2}
\end{eqnarray}
where the set of ${\bm \pi}$ consists of two sub-sets of binary $0,1$ variables defined on the
vertexes of the graph and on the edges of the graph, respectively: ${\bm \pi}\equiv{\bm
\pi}_v\cup{\bm\pi}_e$,  ${\bm \pi}_v\equiv(\pi_a=0,1; a\in{\cal G}_0)$,  and ${\bm
\pi}_e\equiv(\pi_{ab}=0,1; \{a,b\}\in{\cal G}_1)$. The last term on the rhs of Eq.~(\ref{MD2}) stated
in terms of the Kroneker symbols describes the set of the monomer-dimer exclusions. In other words,
a monomer-dimer configuration corresponds to a coloring of the graph (its vertexes and edges) in
such a way that either at least one edge adjusted to the vertex is colored and then the vertex is
not colored, or the adjusted vertexes are all uncolored and then the vertex is colored.

Note that Eq.~(\ref{MD1}) after some obvious modification can be viewed as a definition of an
effective action $S(\bar{{\bm\theta}},{\bm\theta})$ that depends on the fermion variables only
\begin{eqnarray}
\label{MD1-effective} \int_{{\cal G}_1}{\cal
D}{\bm\sigma}e^{S_{0}(\bar{{\bm\theta}},{\bm\theta};{\bm\sigma})}=
e^{S(\bar{{\bm\theta}},{\bm\theta})}, \;\;\; S(\bar{{\bm\theta}},{\bm\theta})= \sum_{a\in{\cal
G}_{0}}w_{a}\bar{\theta}_{a}\theta_{a}+\sum_{\{a,b\}\in{\cal
G}_{1}}w_{ab}\bar{\theta}_{a}\theta_{a}\bar{\theta}_{b}\theta_{b},
\end{eqnarray}
As it usually happens in gauge theories, integration over the gauge field creates fermion
interactions (second term in the action in Eq.~(\ref{MD1-effective})). The interaction can be
decoupled by introducing a Hubbard-Stratonovich field represented by another $\mathbb{Z}_{2}$ gauge
field coupled to $\theta_{a}\theta_{b}$ and $\bar{\theta}_{a}\bar{\theta}_{b}$. This results ina
representation of the partition function of the Monomer-Dimer model in a form of an integral (sum)
over the gauge field, with the integrand represented as a product of two gauge-field dependent
Pfaffians. This representation will be studied in detail in the next paper of the series, with the
focus on its applications to fat graphs.

\subsection{Oriented Disjoint cycle (Determinant) Series}
\label{subsec:DCS}

We further represent the integrand of the GGM partition function (\ref{GGM2}) in the following
simple form
\begin{eqnarray}
\label{DCS1} e^{S_{0}(\bar{{\bm\theta}},{\bm\theta};{\bm\sigma})}=\prod_{a\in{\cal
G}_{0}}e^{w_{a}\bar{\theta}_{a}\theta_{a}}\prod_{\{a,b\}\in{\cal
G}_{1}}\left(e^{H_{ab}\bar{\theta}_{a}\theta_{b}+H_{ba}\bar{\theta}_{b}\theta_{a}}
+(\sigma_{ab}-1)(H_{ab}\bar{\theta}_{a}\theta_{b}+H_{ba}\bar{\theta}_{b}\theta_{a})\right),
\end{eqnarray}
using straightforwardly the Grassman variables anticoagulation relations. Direct integration of
Eq.~(\ref{DCS1}) over the gauge variables implies \begin{eqnarray} \label{DCS2} \int_{{\cal
G}_1}{\cal D}{\bm\sigma}e^{S_{0}(\bar{{\bm\theta}},{\bm\theta};{\bm\sigma})}=\prod_{a\in{\cal
G}_{0}}e^{w_{a}\bar{\theta}_{a}\theta_{a}}\prod_{\{a,b\}\in{\cal
G}_{1}}\left(e^{H_{ab}\bar{\theta}_{a}\theta_{b}+H_{ba}\bar{\theta}_{b}\theta_{a}}
-(H_{ab}\bar{\theta}_{a}\theta_{b}+H_{ba}\bar{\theta}_{b}\theta_{a})\right).
\end{eqnarray}

\begin{figure*}
\centering
\vspace{0.1in}
\includegraphics[width=0.8\textwidth]{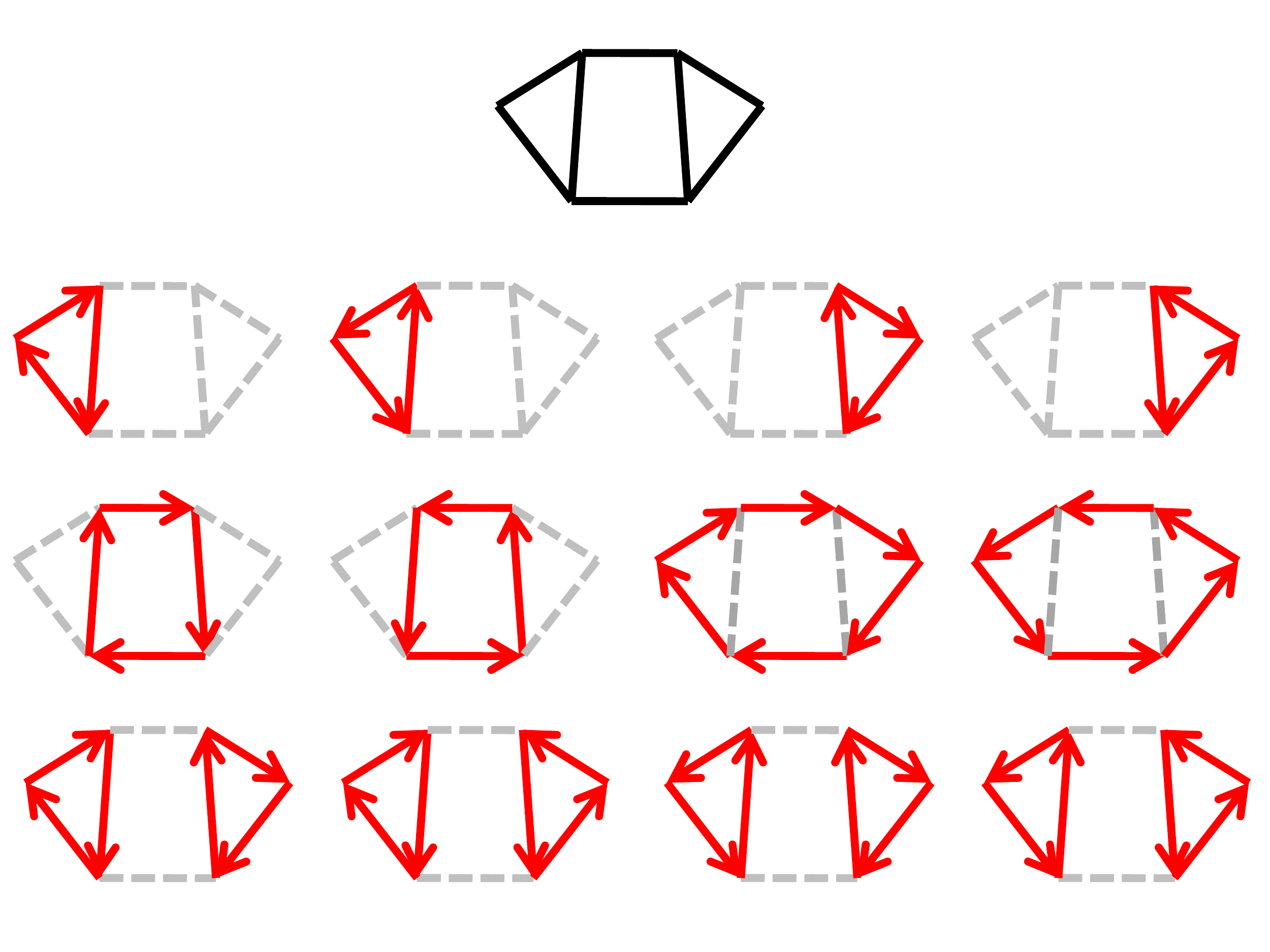}
\caption{Example of a graph (top) and its respective set of oriented disjoint cycles, consisting
of $12$ configurations. Oriented Disjoint cycles are shown in red.} \label{fig:ODC}
\end{figure*}

We further note that Eq.~(\ref{DCS2}) can be represented as a sum of monoms in elements  of $H$.
Let us consider a monom which contains an off-diagonal element $H_{ab}$ but not its conjugate,
$H_{ab}$. Then, it is obvious (from the rules of the Grassman integration) that such a monom  can
only be associated with a directed disjoint cycle which contains the directed segment $(a,b)$,
i.e. the monom should contain a product of the off-diagonal elements along the cycle and do not
contain any of the respective conjugates. Moreover the product of the off-diagonal elements of $H$
along the oriented disjoint cycle originates  primarily from the expansion of the second product
in Eq.~(\ref{DCS2}) in the series.  Therefore, one concludes that Eq.~(\ref{DCS2}) can be
represented as
\begin{eqnarray}
\label{DCS3a} && Z=\sum_{C\in \mbox{ODC}({\cal G})}\bar{r}(C),\quad \bar{r}(C)
=\alpha(C)\det(H|_{{\cal G}\setminus C})\prod_{(a,b)\in C}(-H_{ab}),\\
 && \alpha(C)\equiv\frac{\partial^{|C|}}{\prod_{(a,b)\in C}\partial_{H_{ab}}}\left.
 \left(\int \left(\prod_{a\in C} d\theta_a\right)\left(\prod_{a\in C}d\bar{\theta}_a\right)
 \prod_{(a,b)\in C}\left(e^{H_{ab}\bar{\theta}_{a}\theta_{b}+H_{ba}\bar{\theta}_{b}\theta_{a}}
-(H_{ab}\bar{\theta}_{a}\theta_{b}+H_{ba}\bar{\theta}_{b}\theta_{a})\right)\right)\right|_{H=0}.
\label{DCS3b}
\end{eqnarray}
where $H|_{{\cal G}\setminus C}$ denotes the restriction of $H$ to ${\cal G}\setminus C$. For a
subgraph $C\subset {\cal G}$ we denote by ${\cal G}\setminus C$ the maximal subgraph of ${\cal G}$
that has an empty intersection with $C$. Stated differently, ${\cal G}\setminus C$ is represented
by those edges of ${\cal G}$ that do not have common vertices with $C$. In Eq.~(\ref{DCS3a}) the
$\det$-term corresponds to direct integration over variables that do not belong the oriented
disjoint cycle $C$. In essence, $\alpha(C)$ is a combinatorial factor which is calculated by
straightforward counting. Expanding the integrand in Eq.~(\ref{DCS3b}) into
a series over the square-bracket terms. One finds, that there are $\left(\begin{array}{c} |C|\\
k\end{array}\right)$ contributions associated with a product of $k$ square-bracket terms along the
oriented disjoint cycle, where $|C|$ stands for the length of the oriented disjoint cycle
measured in the number of segments/edges and $1\leq k\leq |C|$, and each of them contributes
$(-1)^{k+1}$ into $\alpha(C)$. Summing up all the nonzero contributions one derives,
\begin{eqnarray}
\alpha(C)=\sum_{k=1}^{|C|}(-1)^{k+1}  \left(\begin{array}{c}  |C|\\ k \end{array}\right)=1.
\label{DCS3c}
\end{eqnarray}
Substituting Eq.~(\ref{DCS3c}) into Eq.~(\ref{DCS3a}) we arrive at the desired expansion of the MD
model partition function with the coefficients represented by determinants
\begin{eqnarray}
\label{DCS-explicit} Z_{{\rm MD}}=\sum_{C\in \mbox{ODC}({\cal G})}\bar{r}(C),\quad \bar{r}(C)
=\det(H|_{{\cal G}\setminus C})\prod_{(a,b)\in C}(-H_{ab})
\end{eqnarray}
An example of a family of oriented disjoint cycles for a sample graph is shown in
Fig.~\ref{fig:ODC}.

\section{Determinant as a Series over Monomer-Dimer Contributions}
\label{sec:DetMD}

Eqs.~(7-10,12) of \cite{08CCa} represent the starting point for discussion of this Section.
However, instead of following the path discussed in \cite{08CCa}, we make another non-BP choice of
the gauge.

The special gauges we will be using are associated with the graph orientations ${\bm\partial} \in
O({\cal G})$, where $O({\cal G})$ denotes the set of graph orientations. An orientation of ${\cal
G}$ associates a direction (``arrow'') on each edge, i.e., it is a pair of maps $\partial_{j}:{\cal
G}_{1}\to{\cal G}_{0}$ with $j=0,1$ so that the edge $\alpha$ connects $\partial_{0}(\alpha)$ and
$\partial_{1}(\alpha)$. For each edge there are two possible orientations:
$\partial_{0}(\{a,b\})=a$, $\partial_{1}(\{a,b\})=b$ and $\partial_{0}(\{a,b\})=b$,
$\partial_{1}(\{a,b\})=a$. In particular ${\rm card}(O({\cal G}))=2^{N_{1}({\cal G})}$, where
$N_{k}={\rm card}({\cal G}_{k})$ with $k=0,1$ represent the number of edges and nodes. Therefore,
orientation can be viewed as a binary variable that resides on the graph edges. The gauge
associated with an orientation ${\bm\partial} \in O({\cal G})$, which is totally determined by
specifying the set $\{\gamma_{ab}({\bm\partial})\}_{a\sim b}$ of numbers that characterize the
local ground states, is given by
\begin{eqnarray}
\label{gauge-orientation} \gamma_{ab}({\bm\partial})&=&(H_{ab})^{-1}\;\; {\rm for} \;
a=\partial_{0}(\{a,b\}), \;\; \gamma_{ab}({\bm\partial})=-(H_{ab})^{-1}\;\; {\rm for} \;
a=\partial_{1}(\{a,b\}), \nonumber \\ \kappa_{ab}&=&1, \;\; c_{ab}=1/2, \;\;
\gamma'_{ab}=-\gamma_{ab}, \;\;\; \zeta_{ab}=(H_{ab})^{-1}
\end{eqnarray}
which simply means that we choose $\gamma_{ab}=\pm (H_{ab})^{-1}$ depending on orientation, and the
signs in front of $H_{ab}$ and $H_{ba}$ are always opposite. The rest of the parameters are
determined by Eq.~(12) of \cite{08CCa}. Note that the set of parameters
$\{\gamma_{ab}({\bm\partial})\}_{a\sim b}$ for a gauge choice given by
Eq.~(\ref{gauge-orientation}) satisfy all the necessary requirements represented by Eq.~(11) of
\cite{08CCa}. Also note that two graph orientations ${\bm\partial}$ and ${\bm\partial}'$ are also
related via a set ${\bm\sigma}$ of edge binary variables: we define $\sigma_{\{a,b\}}=1$ if
$\partial(\{a,b\})=\partial'(\{a,b\})$, and $\sigma_{\{a,b\}}=-1$, otherwise. In particular, a
choice of some base graph orientation allows the graph orientations to be described using the edge
binary variables ${\bm\sigma}$. However, a generic unoriented graph is not equipped with a
preferred choice of orientation.

For any choice of a special gauge (\ref{gauge-orientation}) the Grassman integral
representation (Eq.~(7) of \cite{08CCa}) for the determinant of $H$ can be represented in the
following form
\begin{eqnarray}
\label{HtoMD1} && \det(H)=\left(\prod_{\{a,b\}\in{\cal
G}_1}(-H_{ab}H_{ba})\right)\left(\prod_{a\in{\cal G}_{0}}H_{aa}\right)\int{\cal D}{\bm\chi}{\cal
D}\bar{{\bm\chi}}\prod_{a\in{\cal G}_{0}} e^{(H_{aa})^{-1}\sum_{b\in{\cal
G}_{0}}^{b\sim a}\bar{\chi}_{ba}\sum_{b'\in{\cal G}_{0}}^{b'\sim a}\chi_{b'a}}\nonumber\\
&& \times \prod_{\{a,b\}\in{\cal
G}_{1}}\left(\underbrace{1-\frac{\bar{\chi}_{ab}\chi_{ab}\bar{\chi}_{ba}\chi_{ba}}
{H_{ab}H_{ba}}}_{\mbox{even states at }\{a,b\}}+
\underbrace{\frac{\bar{\chi}_{ab}\chi_{ba}}{H_{ab}}+\frac{\bar{\chi}_{ba}\chi_{ab}}{H_{ba}}}_
{\mbox{odd states at }\{a,b\}}\right),
\end{eqnarray}
which is explicitly independent on the choice of a special gauge. We further partition each factor
on the second line of Eq.~(\ref{HtoMD1}) that corresponds to an edge $\{a,b\}$ into a sum of two
terms, referred to even and, odd according to the terminology introduced (and explained) in
\cite{08CCa}. Our next step is to expand the product of edge terms in the integrand of
Eq.~(\ref{HtoMD1}) into a polynomial over the odd states, followed by performing integration over
the edge Grassman variables that correspond to the odd contributions. For a given choice of the
local odd states we denote by $C\subset {\cal G}$ the subgraph of ${\cal G}$ formed by the edges,
where the odd terms [the third or forth term in the second line of Eq.~(\ref{HtoMD1})] have been
chosen. We start with demonstrating that all vertices of the subgraph $C\subset {\cal G}$ have the
valence two, i.e., $C$ is represented by a disjoint union of simple loops. This follows from the
fact that the expression in the exponent in Eq.~(\ref{HtoMD1}) is actually a product of two linear
combinations of the original Grassman variables and, therefore
\begin{eqnarray}
\label{expand-e}  e^{(H_{aa})^{-1}\sum_{b\in{\cal G}_{0}}^{b\sim a}\bar{\chi}_{ba}\sum_{b'\in{\cal
G}_{0}}^{b'\sim a}\chi_{b'a}}=1+(H_{aa})^{-1}\sum_{b\in{\cal G}_{0}}^{b\sim
a}\bar{\chi}_{ba}\sum_{b'\in{\cal G}_{0}}^{b'\sim a}\chi_{b'a}.
\end{eqnarray}
Consider a vertex $a\in C_{0}$. For the integral over the local vertex variables
$d{\bm\chi}_{a}d\bar{\bm\chi}_{a}$ not to vanish the integrand should contain each of the local
variables $\chi_{ba}$ and $\bar{\chi}_{ba}$ with $b\sim a$ exactly once. These local variables in
the integrand originate from the odd terms, described above, from the even terms [second
contribution in the second line of Eq.~(\ref{HtoMD1})] and from the relevant exponential terms
represented by Eq.~(\ref{expand-e}). If an edge $\{b,a\}\in {\cal G}_{1}$ also belongs to $C$ we have either
the odd term $(H_{ab})^{-1}\bar{\chi}_{ab}\chi_{ba}$ or $(H_{ba})^{-1}\bar{\chi}_{ba}\chi_{ab}$ in
the integrand. Consider the first option (the second option is considered in a similar way), the
local conjugate variable $\bar{\chi}_{ba}$ can originate only from the vertex term given by
Eq.~(\ref{expand-e}) and is represented by a contribution $(H_{aa})^{-1}\bar{\chi}_{ba}\chi_{b'a}$.
The variable $\bar{\chi}_{b'a}$ conjugate to the the variable $\chi_{b'a}$ can originate only from
an odd edge contribution, namely $(H_{b'a})^{-1}\bar{\chi}_{b'a}\chi_{b'a}$, which implies that
$\{b',a\}\in C_{1}$. The edges $\{b,a\}$ and $\{b',a\}$ are the only edges adjacent to the node $a$
that belong to $C_{1}$ since the vertex term [Eq.~(\ref{expand-e})] that provides the conjugate.
local variables contains products of only two Grassmans. Consideration of the other odd
edge term $(H_{ba})^{-1}\bar{\chi}_{ba}\chi_{ab}$ leads to a similar result, but with the opposite
orientation. Therefore, any vertex $a\in C$ has the valence two, and, therefore $C\in
\mbox{ODC}({\cal G})$.

Consider a simple oriented loop $(a_{1},\ldots,a_{n})$, where naturally $a_{j}\sim a_{j+1}$ and
$a_{n}\sim a_{1}$. The associated contribution given by the integral of the loop edge variables of
the product of the edge and vertex contributions has a form
$\prod_{j=1}^{n}(H_{a_{j}a_{j}})^{-1}\prod_{j=1}^{n}(H_{a_{j-1}a_{j}})^{-1}I_{a_{1}\ldots a_{n}}$,
where
\begin{eqnarray}
\label{simple-loop-contr} I_{a_{1}\ldots
a_{n}}&=&\int\prod_{j=1}^{n}d\chi_{a_{j-1}a_{j}}d\bar{\chi}_{a_{j-1}a_{j}}
d\chi_{a_{j+1}a_{j}}d\bar{\chi}_{a_{j+1}a_{j}}\bar{\chi}_{a_{1}a_{2}}\chi_{a_{2}a_{1}}
\bar{\chi}_{a_{2}a_{3}}\chi_{a_{3}a_{2}}\bar{\chi}_{a_{3}a_{4}}\chi_{a_{4}a_{3}}\ldots
\bar{\chi}_{a_{n}a_{1}}\chi_{a_{1}a_{n}} \nonumber \\ &\times&
\prod_{j=1}^{n}\bar{\chi}_{a_{j+1}a_{j}}\chi_{a_{j-1}a_{j}} \nonumber
\\ &=& (-1)^{n-1}\prod_{j=1}^{n}\int d\chi_{a_{j-1}a_{j}}d\bar{\chi}_{a_{j-1}a_{j}}
d\chi_{a_{j+1}a_{j}}d\bar{\chi}_{a_{j+1}a_{j}}\bar{\chi}_{a_{j-1}a_{j}}\chi_{a_{j+1}a_{j}}
\bar{\chi}_{a_{j+1}a_{j}}\chi_{a_{j-1}a_{j}}=-1
\end{eqnarray}
and in Eq.~(\ref{simple-loop-contr}) we use a cyclic convention $j+n=j$. The first equality in
Eq.~(\ref{simple-loop-contr}) is obtained by performing permutations in the following way. We start
with moving the Grassman $\chi_{a_{4}a_{3}}$ in the integrand by two places to the left, followed
by moving the combination $\bar{\chi}_{a_{2}a_{3}}\chi_{a_{4}a_{3}}$ to combine it with the
combination $\bar{\chi}_{a_{4}a_{3}}\chi_{a_{2}a_{3}}$ in the product over $j$, which corresponds
to the value $j=3$; after that we permute the Grassmans $\chi_{a_{3}a_{2}}$ and
$\bar{\chi}_{a_{3}a_{4}}$. The overall permutation provides a $(-1)$ sign factor. Repeating a
similar operation $(n-1)$ times (including the first explicitly described operation) results in the
first equality. The second equality follows from the fact that each of $n$ Grassman integrals in
the intermediate expression is equal to $(-1)$.

Due to Eq.~(\ref{simple-loop-contr}) the resulting expression for the determinant adopts a form
\begin{eqnarray}
&& \det(H)=\sum_{C\in \mbox{ODC}({\cal G})}(-1)^{{\rm deg}(C)} \left(\prod_{\{a,b\}\in
C}(-H_{ab})\right)
Z_1({\cal G}\setminus C),\label{HtoMD2}\\
&& Z_1({\cal G}\setminus C)=\left(\prod_{\{a,b\}\in ({\cal G}_{C})_1}(-H_{ab}H_{ba})\right)
\left(\prod_{a\in ({\cal G}_{C})_{0}}H_{aa}\right)\nonumber\\ && \times \int{\cal
D}{\bm\chi}_{{\cal G}_{C}}{\cal D}\bar{{\bm\chi}}_{{\cal G}_{C}}\prod_{a\in {(\cal G}_{C})_{0}}
\left(1+(H_{aa})^{-1}\sum_{b\in {(\cal G}_{C})_{0}}^{b\sim a}\bar{\chi}_{ba} \sum_{b'\in ({\cal
G}_{C})_{0}}^{b'\sim a}\chi_{b'a}\right)\prod_{\{a,b\}\in ({\cal G}_{C})_{1}}
\left(1-\frac{\bar{\chi}_{ab}\chi_{ab}\bar{\chi}_{ba}\chi_{ba}}{H_{ab}H_{ba}}\right),
\label{HtoMD3}
\end{eqnarray}
where ${\rm deg}(C)$ denotes the number of connected components in $C$. Stated differently, an
element $C\in \mbox{ODC}({\cal G})$ is represented by a disjoint union of oriented simple loops
with ${\rm deg}(C)$ denoting the number of simple loops in $C$. In Eq.~(\ref{HtoMD2})
$\bar{{\bm\chi}}_{{\cal G}_{C}}$, ${\bm\chi}_{{\cal G}_{C}}$ denote the edge Grassman variables
restricted to the subgraph ${\cal G}_{C}$, formed by the edges of ${\cal G}$ that do not belong to
$C$. In deriving Eq.~(\ref{HtoMD3}) we have also made use of Eq.~(\ref{expand-e}) to replace the
exponential vertex terms with their polynomial counterparts.

It is now straightforward to check [by expanding the integrand in Eq.~(\ref{HtoMD3}) into a
polynomial followed by performing integration over the Grassman variables in Eq.~(\ref{HtoMD3})]
that $Z_1({\cal G}')$ is nothing else than the partition function (\ref{MD2}) of the monomer-dimer
model on the graph ${\cal G}$. Consider an edge $\{a,b\}\in({\cal G}_{C})_{1}$. The Grassman
variables $\bar{\chi}_{ab}$, $\bar{\chi}_{ba}$, $\chi_{ab}$ $\chi_{ba}$ whose product provides a
nonzero contribution to the integral over the edge variables can originate from the vertex or edge
terms in Eq.~(\ref{HtoMD3}). If they originate from the edge term, then combining with the
corresponding edge prefactor [from the first line in Eq.~(\ref{HtoMD3})] we obtain the contribution
\begin{eqnarray}
\label{edge-contribution} -H_{ab}H_{ba}\int d\chi_{ab}d\bar{\chi}_{ab}d\chi_{ba}d\bar{\chi}_{ba}
\left(-\frac{\bar{\chi}_{ab}\chi_{ab}\bar{\chi}_{ba}\chi_{ba}}{H_{ab}H_{ba}}\right)=1,
\end{eqnarray}
if they come from the vertex terms associated with the vertices $a$ and $b$, then combined with the
corresponding vertex prefactors, the contribution has a form
\begin{eqnarray}
\label{vertex-contribution} H_{aa}H_{bb}\int d\chi_{ab}d\bar{\chi}_{ab}d\chi_{ba}d\bar{\chi}_{ba}
(H_{bb})^{-1}\bar{\chi}_{ab}\chi_{ab}(H_{aa})^{-1}\bar{\chi}_{ba}\chi_{ba}=1.
\end{eqnarray}
We call such an edge a dimer. Obviously, any node can have not more than one dimer
edge attached to it. The nodes that do not have dimers attached to them are referred to as
monomers. A monomer node $a$ does not provide the Grassmans associated with the vertex term and,
therefore, the prefactor term $H_{aa}$ is not compensated. A dimer $\{a,b\}$ does not provide the
edge terms and, therefore, the edge prefactor $(-H_{ab}H_{ba})$ is not compensated. It is easy to
see that any configuration of monomers and dimers that provides a non-zero contribution to the
Grassman integral in Eq.~(\ref{HtoMD3}) satisfies the monomer-dimer matching rules. Therefore,
$Z_1({\cal G}\setminus C)$ represents the partition function of the monomer-dimer model with the
monomer and dimer weights $w_{a}=H_{aa}$ and $w_{ab}=-H_{ab}H_{ba}$, respectively.

Summarizing,
\begin{eqnarray}
\forall C,\quad {\cal G}'={\cal G}\setminus C:\quad Z_1({\cal G}')=Z_{{\rm MD}}({\cal G}').
\label{HtoMD4}
\end{eqnarray}
which implies

\begin{eqnarray}
\label{LS-SL-explicit-2} \det(H)=\sum_{C\in ODC({\cal G})}r(C), \;\;\; r(C)=(-1)^{{\rm
deg}(C)}\prod_{(a,b)\in C}(-H_{ab}) Z_{{\rm MD}}({\cal G}\setminus C)
\end{eqnarray}

To conclude, we just showed that the determinant of a matrix can be represented in terms of a
series over disjoint oriented cycles of the underlying graph, with each term of the expansion being
proportional to the partition function of the monomer-dimer model defined on the remainder of the
graph, i.e., after the cycles, as well as all edges connected to their vertices are removed.

Comparing Eq.~(\ref{DCS-explicit}) with  Eq.~(\ref{LS-SL-explicit-2}) one finds that in a sense one
is an inverse of the other. While the former expresses the partition function of the MD model on
the graph in terms of an expansion over the determinants (each corresponds to a directed disjoint
cycle), the later does exactly the opposite by expressing the determinant as a series over the
partition functions of the MD models each associated with the exclusion of a directed disjoint
cycle. More details on this relation are given in Appendix \ref{sec:app}.

We complete this Section by  addressing the issue of the gauge invariance of the simple-loop
decomposition. To that end we twist the matrix $H$ as described at the beginning of Section
\ref{subsec:GGM}, i.e. introducing the matrix $H({\bm\sigma})$, twisted by the gauge field
${\bm\sigma}$ as $H_{ab}({\bm\sigma})=\sigma_{ab}H_{ab}$ for $a\ne b$ and
$H_{aa}({\bm\sigma})=H_{aa}$. Applying Eq.~(\ref{LS-SL-explicit-2}) to $H({\bm\sigma})$, recalling
the definition of the $\zeta$-function (\ref{zeta}), and noting that the partition functions
$Z_{{\rm MD}}({\cal G}\setminus C)$ are obviously invariant with respect to the twisting we obtain
the following decomposition for the $\zeta$-function:
\begin{eqnarray}
\label{LS-SL-zeta} \zeta({\bm\sigma})=\det(H({\bm\sigma}))=\sum_{C\in ODC({\cal G})}(-1)^{{\rm
deg}(C)}\prod_{(a,b)\in C}(-\sigma_{ab}H_{ab}) Z_{{\rm MD}}({\cal G}\setminus C)=\sum_{C\in
ODC({\cal G})}r(C)\prod_{(a,b)\in C}\sigma_{ab}
\end{eqnarray}
Therefore, $r(C)$ can be viewed as the coefficients in the expansion of the $\zeta$-function
$\zeta({\bm\sigma})$ in the gauge field ${\bm\sigma}$ and, therefore they do not depend on a
particular way they are evaluated.

\section{Summary and Conclusions}
\label{sec:sum}

To summarize, this manuscript reports new relations between the partition function $Z_{MD}({\cal
G})$ of the monomer-dimer model, defined on an arbitrary graph ${\cal G}$ and the corresponding
determinant of the matrix $H$ and its minors, constructed from the monomer-dimer weights on the
graph. We  have formulated a Graphical Gauge Model (GGM) on a graph, stated in terms of Grassman
variables and binary gauge fields, so that all the relations reported in the paper follow in a
straightforward way via simple and natural manipulations (reparametrizations and integrations) over
the partition function of the GGM. Some results of this paper are also linked to the discussions in
the first paper of the series \cite{08CCa}. In particular, we show here that the expression for a
determinant as an expansion over directed disjoint cycles is related to the Loop Series approach of
\cite{08CCa}. The difference comes from different gauge choices.

In spite of the progress in understanding relations between determinants, loops and matchings (i.e.
valid configurations of the monomer-dimer problems),  there are still many important challenges
left for future analysis. We conclude with mentioning some of these ``natural" challenges.

\begin{itemize}

\item
Given the prominent role the determinants play in the classical studies of the dimer models on
planar graphs and graphs embedded in Riemann surfaces of finite genus
\cite{63Kas,96RZ,99GLa,99GLb,00RZ,07CR,08CR}, one suggests that it should be important to analyze
the consequences of the monomer-dimer, determinant, loops and GGM relations discussed above for
planar and surface graphs, also extending the results of \cite{08CCT}.

\item All the Loop Series related constructions for graphical models, introduced so far in
\cite{06CCa,06CCb,07CC,08CCT,08CCa} and this manuscript, express the partition functions as series
over sub-graphs. On the other hand, the well-known formula $\ln\det(H)={\rm Tr}\ln(H)$, and related
famous expressions for the log-partition function of the Ising model on a planar graph \cite{52KW},
suggests that a multiplicative expansion that represents the partition functions as a product over
sub-graphs, may also exist, at least for some class of graphical models. Exploring possible
multiplicative decompositions constitutes an important theoretical and algorithmic challenge.

\item
One general technical conclusion of the paper is related to the use of Berezin integrals
\cite{87Ber}. Our approach shows that the Grassman-integration technique can be useful for deriving
quantitative exact relations in graphical statistical problems of computer science, operation
research, and information theory. Obviously, the two papers of the series present only the first
step in this direction. A possible extension of this approach, worth a future exploration, would be
to develop a more general super-symmetrical and $\sigma$-models based approach, in the spirit of
\cite{97Efe}, combining normal and Grassman integrations.

\item
To a large extent, the practical utility of the determinant and cycle series discussed in the paper
is yet to be determined. In particular, it remains to be seen weather the reported cycle series
allows an efficient deterministic approximation for the monomer-dimer model partition function. We
speculate that an algorithmic extension of our results may lead to the development of novel Fully
Polynomial-Time Approximation Schemes (FPTAS) for various hard, $\#P$, weighted counting problems
(see e.g. \cite{07BGKNT} for a sample FPTAS example discussed recently).

\end{itemize}

\section{Acknowledgements}

We are grateful to J. Johnson for useful comments. This material is based upon work supported by
the National Science Foundation under CHE-0808910. The work at LANL was carried out under the
auspices of the National Nuclear Security Administration of the U.S. Department of Energy at Los
Alamos National Laboratory under Contract No. DE-AC52-06NA25396.

\bibliographystyle{apsrev}
\bibliography{Gauss,BP_review,Planar}

\appendix
\section{Correspondence between Monomer-Dimer model and Determinant Series}
\label{sec:app}

In this Appendix we establish relation between Eq.~(\ref{DCS-explicit}) and
Eq.~(\ref{LS-SL-explicit-2}) in a somehow straightforward way.

A particular strength of the decomposition of the determinant Eq.~(\ref{LS-SL-explicit-2}) is its
naturality, i.e., it is valid for any graph ${\cal G}$ associated with some matrix $H$. In
particular it can be written for any subgraph ${\cal G}'\subset {\cal G}$. To see the advantages of
naturality in a more clear way we introduce the following notation $X_{C}=\det({\cal G}\setminus
C)$ and $Y_{C}=Z_{{\rm MD}}({\cal G}\setminus C)$ where $C\in ODC({\cal G})$ is an (oriented)
simple loop in ${\cal G}$. For our purposes it is also convenient to introduce an oriented graph
${\cal C}_{\cdot}({\cal G})$ of simple loops, whose nodes are simple loops $C\in ODC({\cal G})$,
i.e., ${\cal C}_{0}({\cal G})=ODC({\cal G})$. The set of links of an oriented graph ${\cal
C}_{1}({\cal G})\subset {\cal C}_{0}({\cal G})\times {\cal C}_{0}({\cal G})$ is naturally a subset
${\cal C}_{1}({\cal G})\subset ODC({\cal G})$ and is defined as follows. We say that $(C,C')$ is an
oriented link (a connecting arrow goes from $C'$ to $C$), i.e. $(C,C')\in {\cal C}_{1}({\cal G})$,
if $C\subset C'$.

The reason why the graph ${\cal C}_{\cdot}={\cal C}_{\cdot}({\cal G})$ has been introduced is that
${\cal C}_{\cdot}$ is the oriented graph associated with the linear relation (matrix) that
expresses the set $\{X_{C}\}_{C\in{\cal C}_{0}}$ of partition functions  in terms of the set
$\{Y_{C}\}_{C\in{\cal C}_{0}}$ of partition functions. To see that we recast
Eq.~(\ref{LS-SL-explicit-2}) for an arbitrary subgraph ${\cal G}'\subset{\cal G}$
\begin{eqnarray}
\label{LS-ODC-explicit} \det({\cal G}')=\sum_{C\in ODC({\cal G}')}(-1)^{{\rm
deg}(C)}\prod_{(a,b)\in C}(-H_{ab}) Z_{{\rm MD}}({\cal G}'\setminus C).
\end{eqnarray}
Applying Eq.~(\ref{LS-ODC-explicit}) for all ${\cal G}'={\cal G}\setminus C$ and making use of the
introduced notation we arrive at
\begin{eqnarray}
\label{det-MD-matrix} X_{C}=\sum_{C'\supset C}R_{CC'}Y_{C'}=\sum_{C'}^{(C,C')\in{\cal
C}_{1}}R_{CC'}Y_{C'}, \;\;\; R_{CC'}=(-1)^{{\rm deg}(C'\setminus C)}\prod_{(a,b)\in C'\setminus
C}(-H_{ab}), \;\;\; R_{CC}=1.
\end{eqnarray}
Note that $R_{CC'}\ne 0$, if and only if $(C,C')\in {\cal C}_{1}$, i.e., $(C,C')$ is an edge of the
oriented graph ${\cal C}_{\cdot}$, which means that ${\cal C}_{\cdot}$ is the oriented graph
associated with the matrix $R_{CC'}$.

The oriented disjoint cycle expansion (\ref{DCS-explicit}) for $Z_{{\rm MD}}$ is obtained by
expressing the inverse matrix $R_{CC'}^{-1}$ as a sum over the oriented (i.e., orientation on the
path should be compatible with the orientation on the graph) paths on the associated graph ${\cal
C_{\cdot}}$:
\begin{eqnarray}
\label{R-C-inverse} Y_{C}=\sum_{C'}R_{CC'}^{-1}X_{C'}, \;\;\; R_{CC'}^{-1}&=&\sum_{p\in P{\cal
C}_{\cdot}}^{p_{0}=C',p_{l(p)}=C}(-1)^{l(p)}\prod_{j=0}^{l(p)-1}R_{p_{j+1}p_{j}} \nonumber
\\ &=&\prod_{(a,b)\in C'\setminus C}(-H_{ab})\sum_{p\in P{\cal
C}_{\cdot}}^{p_{0}=C',p_{l(p)}=C}(-1)^{l(p)+{\rm deg}(C')-{\rm deg}(C)}=\prod_{(a,b)\in C'\setminus
C}(-H_{ab}).
\end{eqnarray}
In deriving Eq.~(\ref{R-C-inverse}) we made use of the fact $R_{CC}=1$ for the diagonal elements
and the expression for the off-diagonal components (\ref{det-MD-matrix}). In particular the
specific form of $R_{CC'}$ implies that the contributions of different paths are the same up to a
sign. The last equality in Eq.~(\ref{R-C-inverse}) is obtained by an explicit computation of the
combinatorial factor
\begin{eqnarray}
\label{comb-factor} \sum_{l=0}^{{\rm deg}(C')-{\rm deg}(C)}(-1)^{l(p)+{\rm deg}(C')-{\rm
deg}(C)}{\cal N}\left(l;{\rm deg}(C')-{\rm deg}(C)\right)=1
\end{eqnarray}
where ${\cal N}(l;N)$ is the number of ways one can put $N$ objects into $l$ boxes with each box
containing at least one object.

Applying Eq.~(\ref{comb-factor}) to $C=\emptyset$ and recalling the meaning of the notation $X_{C}$
and $Y_{C}$ we arrive at Eq.~(\ref{DCS-explicit}).

\section{Expansion of a Determinant and Summation over the Gauges}
\label{sec:app-B}

In this Appendix we present an alternative derivation of the decomposition
(\ref{LS-SL-explicit-2}) of a determinant into a sum over the oriented disjoint cycles with
the individual contributions expressed in terms of the partition functions of the Monomer-Dimer
(MD) models defined on the proper subgraphs of ${\cal G}$.

First of all we note that the loop decomposition (Eq.~(22) of \cite{08CCa}) is valid in any gauge,
i.e., for any choice of the set $\{\gamma_{ab}\}_{a\sim b}$, provided that the summation over
generalized loops $C\in GL({\cal G})$ is extended to the summation over all subgraphs ${\cal
G}'\subset{\cal G}$. In the case of a BP gauge the latter summation is restricted to the summation
over the generalized loops, since the BP gauge ensures the vanishing of the rest of the
contributions. Multiplying the relative contributions $r(C,C')$ with the prefactor and changing the
order of the summations we recast the loop decomposition in a form
\begin{eqnarray}
\label{LS-modified} \det(H)=\sum_{C\in ODC({\cal G})}r(C), \;\;\; r(C)=\sum_{{\cal G}'\subset{\cal
G}}^{C\subset{\cal G}'}Z({\cal G}',C),
\end{eqnarray}
of a decomposition in oriented disjoint cycles. Note that in this notation the BP contribution
corresponds to the empty simple loop and empty subgraph. Note that strictly speaking the loop
series depends on the gauge choice. However, the gauge freedom (among the gauges we are dealing
with) belongs to the boson subspace, which implies that the coefficients $r(C)$ in
Eq.~(\ref{LS-modified}) should be gauge invariant. This issue is addressed at the end of section
\ref{sec:DetMD}.

We will consider the loop series (\ref{LS-modified}) for all $2^{N_{1}}$ special gauges associated
with the graph orientations (they are given by Eq.~(\ref{gauge-orientation})) and average it with
an equal weight of $2^{-N_{1}}$. This is a legitimate procedure since the sum of all terms in a
loop series is naturally gauge invariant. We also note that for given $C\in ODC({\cal G})$ a
particular choice of a subgraph $C\subset {\cal G}' \subset {\cal G}$ can be described by a
particular configuration of a set ${\bm\sigma}\in M_{C}$ of binary variables that reside on those
edges of ${\cal G}$ that do not belong to $C$. Namely, $\sigma_{\alpha}=-1$ for $\alpha\in{\cal
G}'$ (painted edge that correspond to local even excited state) and $\sigma_{\alpha}=1$ otherwise
(local ground state). Combining these arguments with the expressions for the ingredients of the
loop expansion )Eqs.~(23) and (24) in \cite{08CCa}) we arrive at
\begin{eqnarray}
\label{LS-auge-averaged} Z&=&2^{-N_{1}({\cal G})}\sum_{{\bm\partial}\in O({\cal G})}\sum_{C\in
ODC({\cal G})}\sum_{{\cal G}'\subset{\cal G}}^{C\subset{\cal G}'}Z({\cal G}',C;{\bm\partial})
\nonumber
\\ &=&2^{-N_{1}({\cal G})}\sum_{C\in ODC({\cal G})}\sum_{{\bm\partial}\in O({\cal G})}\sum_{{\bm\sigma}\in
M_{C}}(-1)^{{\rm deg}(C)}2^{-(N_{1}({\cal G})-N_{1}(C))}\prod_{(c,d)\in C}H_{cd}\prod_{a\in({\cal
G}\setminus
C)_{0}}\left(-H_{aa}-\sum_{b\sim a}(\gamma_{ba}({\bm\partial}))^{-1}\sigma_{ba}\right) \nonumber \\
&=&2^{-N_{1}({\cal G})}\sum_{C\in ODC({\cal G})}(-1)^{{\rm deg}(C)}2^{-(N_{1}({\cal G})-N_{1}(C))}
\nonumber
\\ &\times& \prod_{(c,d)\in C}H_{cd}\sum_{{\bm\sigma}\in M_{C}}\sum_{{\bm\partial}\in O({\cal
G})}\prod_{a\in({\cal G}\setminus C)_{0}}\left(-H_{aa}-\sum_{b\sim
a}(\gamma_{ba}({\bm\partial}(\{a,b\})))^{-1}\sigma_{ba}\right).
\end{eqnarray}
Comparing Eq.~(\ref{LS-auge-averaged}) with Eqs.~(\ref{LS-modified}) we see that the decomposition
in simple loops (\ref{LS-SL-explicit-2}) is reproduced if we define
\begin{eqnarray}
\label{Z-MD-from-LS} Z_{{\rm MD}}({\cal G}\setminus C)=2^{-(N_{1}({\cal
G})-N_{1}(C))}\sum_{{\bm\sigma}\in M_{C}}2^{-N_{1}({\cal G})}\sum_{{\bm\partial}\in O({\cal
G})}\prod_{a\in({\cal G}\setminus C)_{0}}\left(-H_{aa}-\sum_{b\sim
a}(\gamma_{ba}({\bm\partial}(\{a,b\})))^{-1}\sigma_{ba}\right).
\end{eqnarray}
The only thing we need to show at this point is that the expression in Eq.~(\ref{Z-MD-from-LS})
reproduces the partition function of the MD model on the graph ${\cal G}\setminus C$. This is
achieved by performing the summation over the binary variables ${\bm\partial}\in O({\cal G})$. The
desired result follows from an obvious property
\begin{eqnarray}
\label{sum-orient-local}
\frac{1}{2}\sum_{{\bm\partial}(\{a,b\})}\left(\gamma_{ba}({\bm\partial}(\{a,b\}))\right)^{-1}\sigma_{ba}=0,
\;\;\;
\frac{1}{2}\sum_{{\bm\partial}(\{a,b\})}\left(\gamma_{ba}({\bm\partial}(\{a,b\}))\right)^{-1}
\left(\gamma_{ab}({\bm\partial}(\{a,b\}))\right)^{-1}\sigma_{ba}\sigma_{ab}=-H_{ab}H_{ba},
\end{eqnarray}
where both sums in Eq.~(\ref{sum-orient-local}) contain two terms that correspond to two possible
values of the orientation ${\bm\partial}(\{a,b\})$ of the edge $\{a,b\}$. A choice of a diagonal
term in the parenthesis in Eq.~(\ref{Z-MD-from-LS}) corresponds to having a monomer on the node $a$
with the weight $H_{aa}$. It follows from Eq.~(\ref{sum-orient-local}) that the off-diagonal terms
should always go in pairs, each pair $(\gamma_{ab}\gamma_{ba})^{-1}$ corresponds to having a dimer
on the link $\{a,b\}$, whose weight is $-H_{ab}H_{ba}$. It also follows from
Eq.~(\ref{sum-orient-local}) that the sum over orientations in Eq.~(\ref{Z-MD-from-LS}) does not
depend on ${\bm\sigma}$ and, therefore, the sum over ${\bm\sigma}$ just cancels out the first
prefactor in Eq.~(\ref{Z-MD-from-LS}). This completes the proof.

\end{document}